\catcode`\@=11
\def\gsim{\ifmmode{\mathrel{\mathpalette\@versim>}}
    \else{$\mathrel{\mathpalette\@versim>}$}\fi}
\def\lsim{\ifmmode{\mathrel{\mathpalette\@versim<}}
    \else{$\mathrel{\mathpalette\@versim<}$}\fi}
\def\@versim#1#2{\lower 2.9truept \vbox{\baselineskip 0pt \lineskip 
    0.5truept \ialign{$\m@th#1\hfil##\hfil$\crcr#2\crcr\sim\crcr}}}
\catcode`\@=12
\def\brem{bremsstrahlung$\;\,$}

\def\eps{\epsilon}

\def\lb{\hbox{$L_{\rm B}$}}
\def\lbh{\hbox{$L_{\rm BH}$}}

\def\lsol{\hbox{$L_\odot$}}
\def\lx{\hbox{$L_{\rm X}$}}

\def\mast{M_*}
\def\mgas{\hbox{$M_{\rm gas}$}}
\def\mh{\hbox{$M_{\rm h}$}}
\def\mbh{M_{\rm BH}}
\def\mdot{\dot\mbh}

\def\msol{\hbox{$M_\odot$}}

\def\nub{\nu _{\rm b}}

\def\rch{r_{\rm ch}}

\def\rcs{r_{\rm c*}}

\def\ref{\par\noindent\hangindent 15 pt}

\def\ssc{\sigma_{\rm \circ *}}

\def\t15{t_{15}}

\def\SX{\Sigma_{\rm X}}
\documentstyle[12pt,aasms4]{article}
\slugcomment{In press on ApJ (Letters)}
\lefthead{Ciotti and Ostriker}
\righthead{Cooling flows and QSOs}

\begin{document}

\title{Cooling flows and quasars:}
\title{Different aspects of the same phenomenon? Concepts}

\author{Luca Ciotti\altaffilmark{1} and Jeremiah P. Ostriker}
\affil{Princeton University Observatory, Peyton Hall,
       Princeton, NJ 08544 USA}

% Notice that each of these authors has alternate affiliations, which
% are identified by the \altaffilmark after each name.  The actual alternate
% affiliation information is typeset in footnotes at the bottom of the
% first page, and the text itself is specified in \altaffiltext commands.
% There is a separate \altaffiltext for each alternate affiliation
% indicated above.

\altaffiltext{1}{On leave from Osservatorio Astronomico di Bologna, 
via Zamboni 33, 40126 Bologna, Italy.} 

% The abstract environment prints out the receipt and acceptance dates
% if they are relevant for the journal style.  For the aasms style, they
% will print out as horizontal rules for the editorial staff to type
% on, so long as the author does not include \received and \accepted
% commands.  This should not be done, since \received and \accepted dates
% are not known to the the author.

\begin{abstract}

We present a new class of solutions for the gas flows in elliptical
galaxies containing massive central black holes (BH).  Modified King
model galaxies are assumed.  Two source terms operate: mass loss from
evolving stars, and a secularly declining heating by supernovae
(SNIa).  Relevant atomic physical processes are modeled in detail.
Like the previous models investigated by Ciotti et al. (1991, CDPR),
these new models first evolve through three consecutive evolutionary
stages: wind, outflow, and inflow.  At this point the presence of the
BH alters dramatically the subsequent evolution, because the energy
emitted by the BH can heat the surrounding gas to above virial
temperatures, causing the formation of a hot expanding central bubble.
Short and strong nuclear bursts of radiation ($\lbh$) are followed by
longer periods during which the X-ray galaxy emission comes from the
coronal gas ($\lx$).  The range and approximate distribution spanned
by $\lx$ are found to be in accordance with observations of X-ray
early type galaxies.  Moreover, although high accretion rates occur
during bursting phases when the central BH has a luminosity
characteristic of quasars, the total mass accreted is very small when
compared to that predicted by stationary cooling-flow solutions and
computed masses are in accord with putative BH nuclear masses.  In the
bursting phases the X-ray gas luminosity is low and the surface
brightness profile is very low compared to pre-burst or to cooling
flow models.  We propose that these new models, while solving some
long-standing problems of the cooling flow scenario, can provide a
unified description of QSO-like objects and X-ray emitting elliptical
galaxies, these being the same objects observed at two different
evolutionary phases.

\end{abstract}

% The different journals have different requirements for keywords.  The
% keywords.apj file, found on aas.org in the pubs/aastex-misc directory, 
% contains a list of keywords used with the ApJ and Letters.  These are 
% usually assigned by the editor, but authors may include them in their 
% manuscripts if they wish. 

%\keywords{galaxies: active --- galaxies: cooling flows --- galaxies: evolution
%--- X-rays: galaxies}


% That's it for the front matter.  On to the main body of the paper.
% We'll only put in tutorial remarks at the beginning of each section
% so you can see entire sections together.
% In the first two sections, you should notice the use of the LaTeX \cite
% command to identify citations.  The citations are tied to the
% reference list via symbolic KEYs.  We have chosen the first three
% characters of the first author's name plus the last two numeral of the
% year of publication.  The corresponding reference has a \bibitem
% command in the reference list below.
%
% Please see the AASTeX manual for a more complete discussion on how to make
% \cite-\bibitem work for you.   

\section{Introduction}

% Authors may indicate to the editorial staff where they would like 
% figures and tables to be placed in the manuscript.  This is done with
% either the \placefigure{KEY} or \placetable{KEY} commands.  These
% commands require \label{KEY} commands to be placed appropriately with
% corresponding table and figure captions.  When the manuscript is
% printed a short note is printed on the page where the figure or table
% is to go.  These commands are ignored in the aaspp4 and aas2pp4 styles.
% The \notetoeditor{TEXT} command allows the author to communicate some
% information to the copy editor.  This information will appear as a 
% footnote on the printed copy for the aasms4 style file.  Nothing will 
% appear on the printed copy if the aaspp4 or aas2pp4 style file is used.

As first revealed by {\it Einstein} observations, normal elliptical
galaxies, both isolated or in groups and clusters, can be powerful
X-ray sources with 0.5 - 4.5 KeV luminosities $\lx$ ranging from $\sim
10^{39}$ to $\sim 10^{42}$ erg s$^{-1}$.  This emission is associated
with hot gaseous halos within the galaxies, containing $\mgas=10^8 -
10^{11}\msol$ (see
\cite{fab89}).

In order to explain this observational finding, a certain class of
solutions designated cooling flow models have been proposed and
extensively investigated (e.g., Fabian, Nulsen, \& Canizares 1984;
Sarazin \& White 1987,1988; Vedder, Trester, \& Canizares 1988).
While these models have many attractive features, they are far from
giving a totally satisfactory account of the X-ray properties of all
elliptical galaxies, as most observed systems are much fainter in the
X-rays than the models predict and have different radial profiles than
expected.  Moreover, the cooling flow models do not solve the question
of {\it where} the cool gas is deposited: over a Hubble time an amount
of material comparable to the mass of stars in the galactic core flows
into the nucleus, but the expected distortions of the central
optical surface brightness and velocity dispersion are not observed.

One possible solution to part of the previous set of problems was
proposed by D'Ercole et al. (1989) and CDPR, who showed that the
heating from SNIa could be effective in maintaining low luminosity
galaxies in a wind phase over an Hubble time (and so preventing the
gas from accumulating in the centre).  But the most massive galaxies
ultimately experience a central cooling catastrophe, leading to a
situation similar to a cooling flow.  Clearly a component of the
explanation is missing and is possibly related to the fact (e.g.,
\cite{r84}) that many (perhaps most) early-type galaxies show a
nuclear activity, and, according to the standard interpretation of the
AGN phenomenon, a massive BH is at its origin.  So it is natural to
investigate the accretion of a galactic gas inflow onto galaxies
within which lurk massive central BHs ($\mbh\sim 10^8\msol$).

Binney \& Tabor (1995, BT) explored this problem with the aid of
spherically symmetric numerical simulations, assuming an homogeneous
release of energy in the inner kpc of their galaxy models during the
inflow phases. Moreover, BT assumed that all the accretion luminosity
was available for the gas heating, due to the interaction between a
nuclear jet and the surrounding ISM.

In the present paper we explore, by numerical integration of the fully
non-stationary equations of hydrodynamics, the modifications on the
results of CDPR, assuming the presence of a massive BH the galaxy
centre with detailed allowance for the effects on the flow of the
radiation emitted by the central BH.  As will be shown, the gas over
the body of the galaxy is (as noted by BT) really optically thin, but
nevertheless the effect of {\it energy} exchange between the nuclear
{\it radiation} and the gas flow is dramatic. This effect was already
known and extensively studied for accreting compact objects
(\cite{omccwy76}; Cowie, Ostriker, \& Stark 1978).  In a successive
paper (\cite{co97}, Paper II), a quantitative analysis of various
aspects of the scenario summarized in this Letter, together with an
exaustive description of the input physics and its modelization, will
be given.

\section{Results}

All the results shown here refer to a model whose parameters are fixed
following the line of CDPR.  The stellar density profile is a King
(1972) distribution, with total blue luminosity $\lb=5\times
10^{10}\lsol$, central velocity dispersion $\ssc=280$ km s$^{-1}$, and
core radius $\rcs=350$ pc.  The dark-matter halo is described by a
quasi-isothermal density distribution, with $\mh/\mast=7.8$ and
$\rch/\rcs =4.2$. The SNIa rate is the same as that in the King
Reference Model of CDPR.  The bolometric luminosity emitted by
accretion onto the BH is $\lbh\equiv\eps c^2 \mdot$, where $c$ is the
light velocity, and $\eps$ is the accretion efficiency, with
$10^{-3}\lsim
\eps\lsim 10^{-1}$. The spectral distribution of $\lbh$ near the BH
is assumed to be $\lbh(\nu)\propto
\lbh\nu^{-0.5}/(\nub^{0.7}+\nu^{0.7})$, where $h\nub$=1MeV. In the
present model $\eps=0.1$.

%\placefigure{fig1}

The spherically symmetric hydrodynamical equations are integrated
numerically using the Eulerian up-wind scheme with time splitting and
artifical viscosity as used in CDPR. In the energy equation the
contribution of Compton heating (and cooling) of the gas due to $\lbh$
and to the recycling of the \brem radiation produced by the gas heated
by the BH activity are included. We allow also for the effect of the
photoionization in both cooling and heating of cold gas as well as
momentum exchange between photons and electrons.  At each radius the
radiation field is integrated, considering the gas differential
absorption on $\lbh (\nu)$ and using for the electrons the
Klein-Nishina cross-section. In this way the absorbed fraction of
$\lbh$ is computed self-consistently.

%\placefigure{fig2}

In Fig. 1b (solid line) the temporal evolution of the coronal X-ray
luminosity $\lx$ of the gas in the 0.5--4.5 KeV band is shown over an
Hubble time.  The evolution up to the so-called {\it cooling
catastrophe} ($t\simeq 9.4$ Gyr) is analogous to that described in
CDPR, but after this time the Compton heating instability completely
alters the flow evolution and its properties.  At the cooling
catastrophe negative infall velocities appear near the galaxy center,
with $\mdot\sim 60\;\msol$yr$^{-1}$, and this accretion produces a
strong, energetic feed-back producing a very high $\lbh$ (Fig. 1a).
The gas in the central regions of the galaxy is strongly heated to
temperatures comparable with the Compton temperature associated with
$\lbh (\nu)$ ($\simeq 10^9$ K), and starts to expand, decreasing its
density by more than two orders of magnitude, driving a shock wave
outwards and producing a hot bubble of a few hundred parsecs in
diameter. The net effect is, observationally, a large reduction of
$\lx$ (Fig. 1b), and, hydrodynamically, the interruption of the
galactic inflow and the consequent shut-off of $\lbh$.  Then the
radiative losses increase again, and, after a period of the order of
the hot gas cooling time, the cycle repeats.  In the model described
here this time is of the order of $\sim 1$ Gyr.  In the case of very
high accretion the shock wave can reach the galaxy edge, and expel gas
from the galaxy.  At higher time resolution each burst shows a very
complex structure, that will be discussed in detail in Paper II: the
temporal blow-up of the first burst shown in Fig. 1 is plotted in
Fig. 2, showing QSO-like luminosities.  An important characteristic of
all computed models -- of which a single representative is here
discussed -- is that the fraction of $\lbh$ effectively absorbed by
the gaseous halo is in the range $10^{-4}-10^{-2}$, but {\it the gas
flows are found to be invariably unstable due to Compton heating for
all the explored efficiencies}: in presence of a massive BH at the
center of elliptical galaxies the possibility of a stationary cooling
flow seems to be very remote.

%\placefigure{fig3}

In Fig. 3 we show the distribution of $\lx$ from 9 Gyr to 15 Gyr.  The
dashed histogram shows the model distribution of $\lx$ given in
Fig. 1, and the solid line shows data for (non-boxy) early-type
galaxies taken from Fig.1 of CDPR.  Finally the dotted histogram is
the distribution of $\lx$ for the same model, with the cooling flow
assumption of $\eps=0$.  We see that the model with $\eps =0.1$ has a
distribution, over time, of $\lx$ surprisingly similar to that of
observed galaxies, but the cooling flow model (as is well known)
produces far too much radiation. In Fig. 4 the X-ray surface
brightness profile ($\SX$) of the presented model is shown at two
different epochs, before and during bursts (vertical arrows in
Fig. 2). Also shown is the cooling flow profile for the same galaxy at
$t=15$ Gyr.  Note how $\SX$ is characterized by a well defined core
before a burst, alleviating the problem of the too cuspy $\SX$ that
afflicts cooling flow models (Canizares, Fabbiano, \& Trinchieri
1987).  Certainly interesting is the fate of the transient cold shell
surrounding the hot bubble (especially in low-$\eps$ solutions) during
the flaring activity, when the central gas surface brightness is very
low.  Due to Rayleigh-Taylor instability, the shell will break up, and
perhaps cold fingers of gas should be observable inside the hot low
density bubble, accreting on the central BH. So, the presence of the
central heating source produces in a natural way a multiphase ISM on
galactic scales, while the same phenomenon may be harder to obtain
(Balbus 1991) in the cooling flow scenario.

Fig. 5 shows the \brem spectra in the preburst (dotted) and during
burst (solid) phases, compared to the cooling flow (dashed) spectrum.
The emitted spectrum is never as soft in this set of models as it is
in cooling flow models, and, during bursts, occasionally it will have
a very hard tail.

%\placefigure{fig4}

\section{Discussion and Conclusions}

In this Letter we show how the presence of a massive central BH in
early type galaxies is able to produce naturally both the observed
X-ray underluminosity with respect to the pure cooling flow
expectations, and the large observed scatter in $\lx$ at fixed $\lb$.
As can be seen from Fig. 1, $\lx$ -- except in the very short period
of bursts -- is always lower than that of the corresponding inflow
model with $\eps=0$. Moreover, the statistical distribution of
observed data compared with the amount of time spent at each $\lx$ by
the model here discussed, is eloquent (Fig. 3). Finally, due to the
strong feed-back on the gas flows of the radiation emitted by the
accretion, the total mass accumulated by the BH over 15 Gyr is very
low ($\sim 3\;10^8\msol$), to be compared with the $\sim 10^{10}\msol$
of the correspondent $\eps=0$-model. The same model in pure
cooling-flow (without the initial SNIa driven wind-outflow phases)
would have accumulated in its center $\sim 10^{11}\msol$ of gas.

From an observational point of view, it is interesting to note that
during the accretion phases the galaxy luminosity is dominated by
$\lbh$ (with highest values at $10^{46}-10^{47}$ erg s$^{-1}$), while
during the quiescent BH phases the galaxy emission is due only to the
diffuse hot gas $\lx$.  The total energy emitted by the accretion when
$\lbh >10^{42}$ is $\sim 7.5\,10^{61}$ erg, while during the same
phases the total energy emitted by the coronal gas is $\sim
5.6\,10^{57}$ erg.  The ratio between the total time spent by the
galaxy when $\lbh >\lx$, and the total time spanned by the simulation
is $\sim 10^{-2}$: very few galaxies should be caught in a AGN-like
phase even though most contain central BHs.

%\placefigure{fig5}

Thus, the Compton heating instability could be an alternative
possibility to that advocated by Fabian \& Rees (1995) in order to
explain why the nuclei of elliptical galaxies are not luminous sources
of radiation as expected if they host a massive central BH.  A clear
prediction of this model is that some significant fraction of QSOs
should be embedded in high temperature, low surface brightness X-ray
\brem halos.

In other works to be reported in Paper II we varied the efficiency in
the range $10^{-3}\leq\eps\leq 10^{-1}$ and the galaxy luminosity in
the range $10^{10}\leq\lb/\lsol\leq 10^{11}$, with results very
similar to those shown in Figs. 2-5.  We are well aware that in the
real accretion phenomenon a disk geometry for the infalling gas seems
to be inescapable. Then the accretion luminosity will be emitted
preferentially along polar directions. It is clear that (at least)
fully 2D hydrodynamical simulations are required for a better
understanding of this problem, and to follow the development and the
final fate of the gas instabilities.  We are working in this
direction, but expect that many of the quantitative features of the
present work will be carried over to the more complicated
calculations.

\acknowledgments

We would like to thank Giuseppe Bertin, James Binney, Annibale
D'Ercole, Bruce Draine, Silvia Pellegrini, and Alvio Renzini for
useful discussions and advice.  This work was supported by NSF grant
AST9424416 and in addition L.C. was supported by NSF grant AST9108103
and by the Italian Space Agency (ASI) with contracts ASI-94-RS-96 and
ASI-95-RS-152.

% And finally, we must deal with the figures.  There are three figures
% associated with this manuscript; two figures are Encapsulated
% PostScript (EPS) files.  The third figure is a grey scale figure that does
% not exist in EPS form.
%
% Authors have three options for including figure information within a 
% manuscript.  Not all the options may be acceptable by the target Journal - be
% sure to look at the appropriate submission instructions, electronic or 
% otherwise.
%
% Option 1.  Using this option, only the figure captions are included in the
% main body of the manuscript.  The figure captions must start on a new page.
% The captions are generated with the \figcaption[]{} command: the first 
% argument is optional, if you put something in there, put the name of the 
% EPS file that goes with the caption; the second argument is the figure 
% caption itself, and may include a \label command.  The \figcaption command
% generates the figure numbers.  This option is acceptable for all manuscript
% submissions.

\clearpage

\figcaption[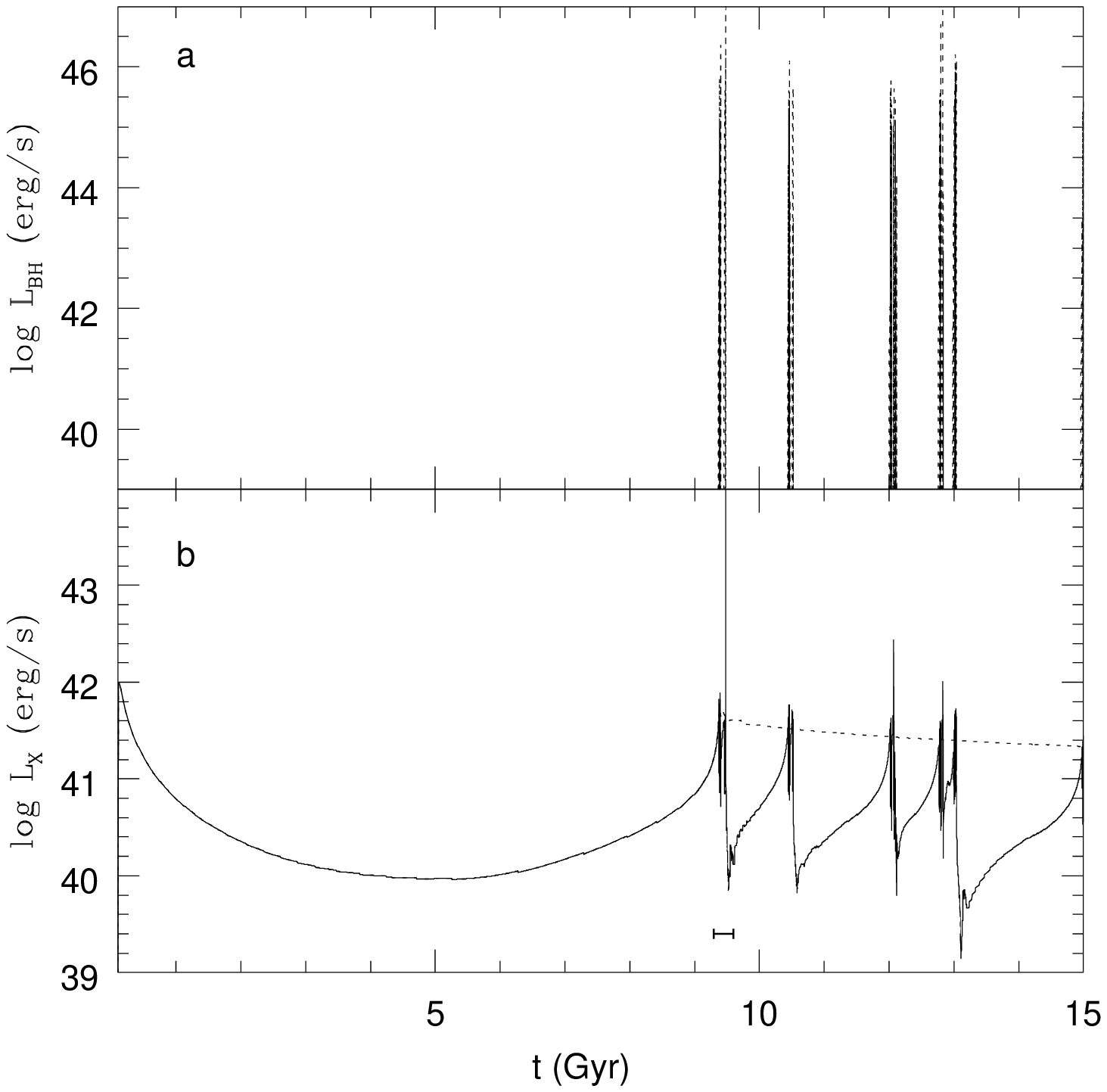]
           {Panel (a): the time evolution of $\lbh$ (bolometric) emitted
           at the galaxy centre. Panel (b): the time evolution of $\lx$ for  
           the model with $\eps=0.1$ (solid line), and that of the same model 
           with $\eps=0$ (cooling flow - dashed line). $\lx$ is calculated 
           inside the galaxy truncation radius and in the range 0.5--4.5 KeV.
           Time interval in horizontal error bar is expanded in Fig. 2.  
           \label{fig1}}

\figcaption[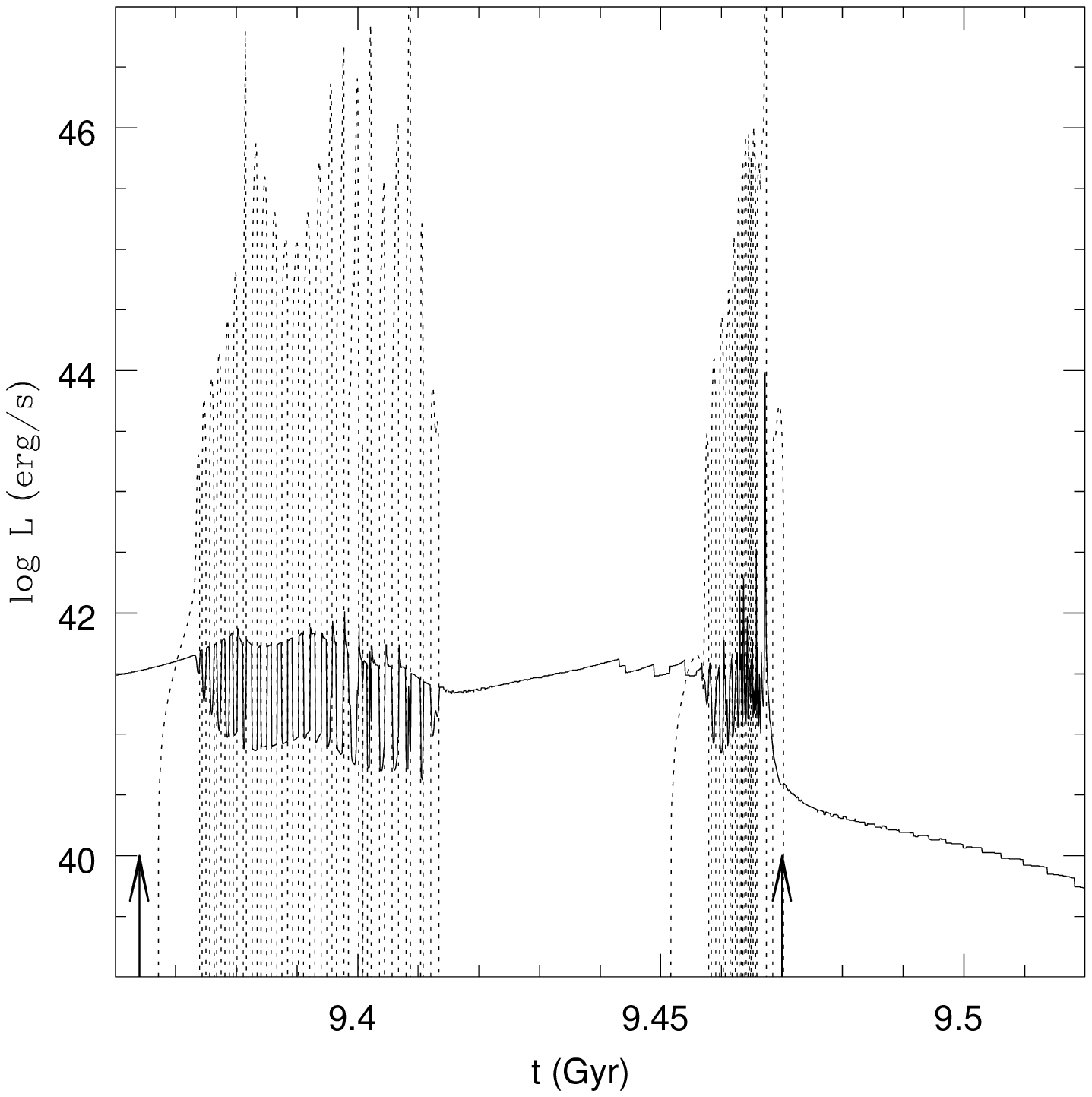]
           {Time expansion of the first burst shown in Fig. 1. The solid line 
           is $\lx$, the dotted line $\lbh$. The temporal sub-structure of 
           the burst is apparent, and the quasar-like luminosity $10^{45}<\lbh
           /{\rm erg}\, {\rm s}^{-1}<10^{47}$ is seen during bursts.
           Arrows mark epochs of ``before'' and ``during'' bursts referred
           to in text. \label{fig2}}

\figcaption[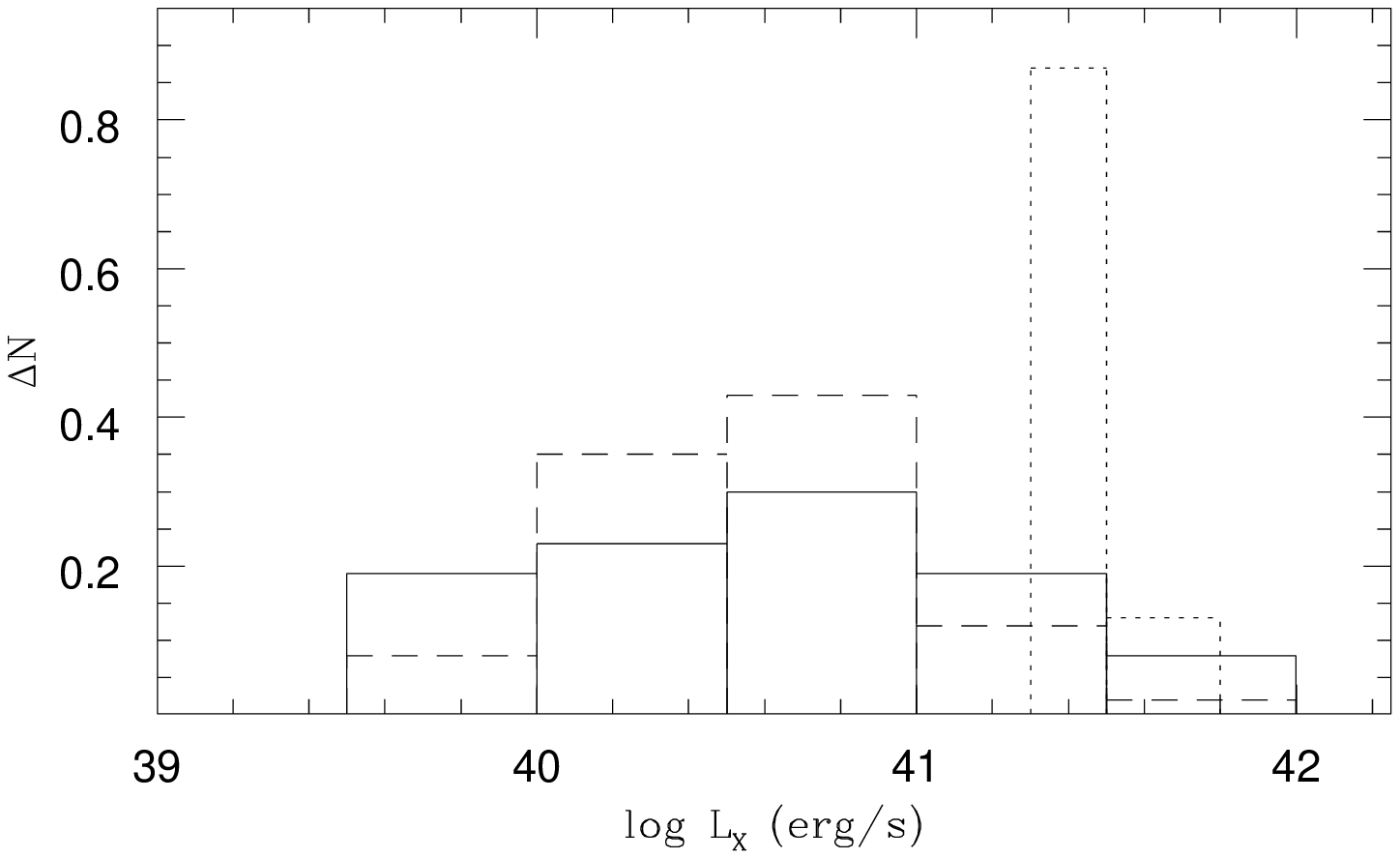]
           {The statistical distribution of $\lx$ for observed galaxies
           (solid) in the range $10.4<\log(\lb/\lsol)<10.8$ derived from 
           Fig. 1 of CDPR. The dashed histogram represents the time 
           distribution of $\lx$ for the presented model from 9 Gyr to 15 Gyr,
           while the dotted histogram shows the cooling flow ($\eps=0$) 
           model; clearly the bursting model provides a better fit to the 
           observed distribution of $\lx$. \label{fig3}}

\figcaption[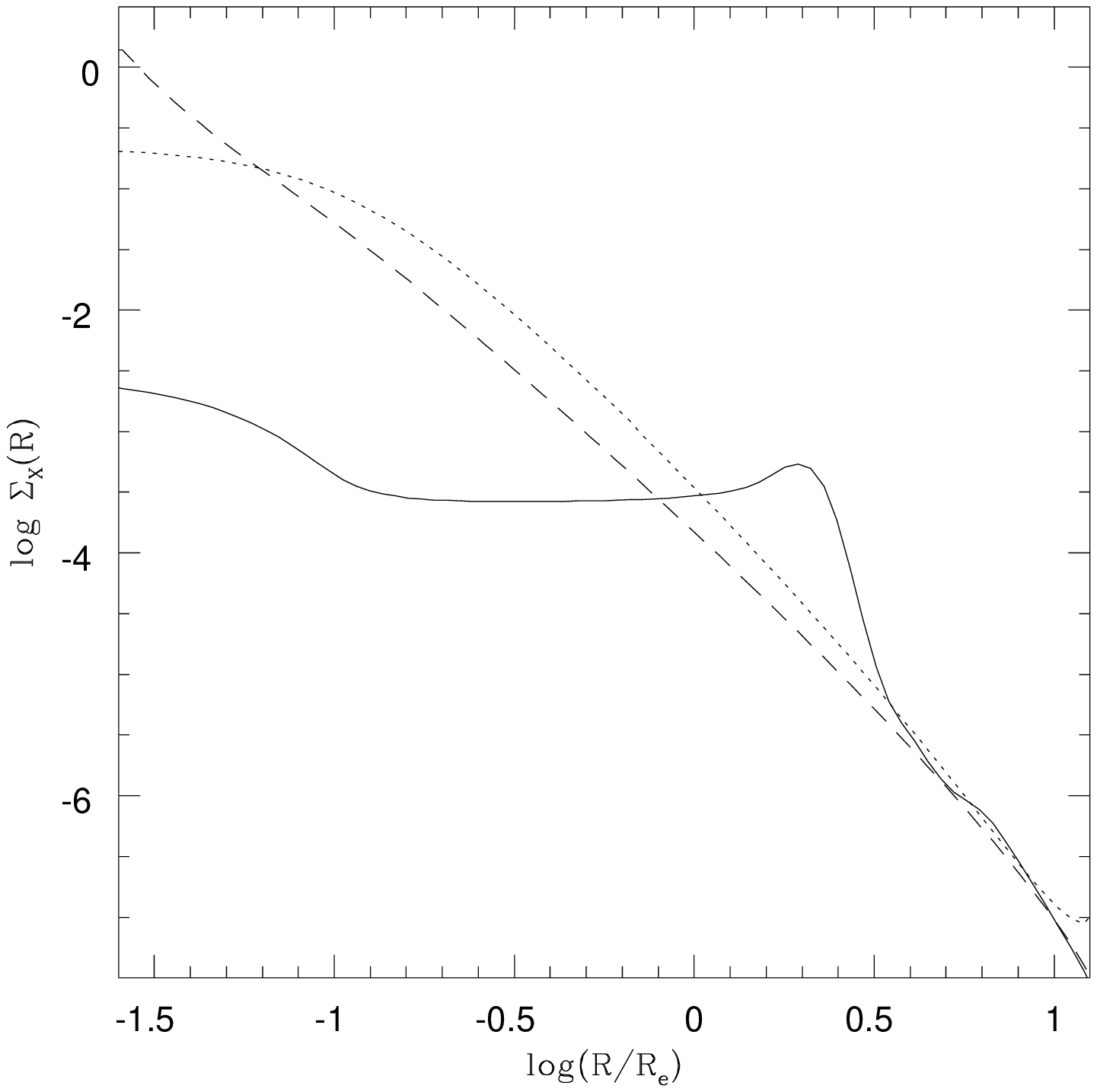]
            {The surface brightness $\SX$ profile (in arbitrary units)
            at $t\simeq 9.36$ (dotted line, immediately before a burst), 
            and $t=9.47$ (solid line, during bursts), corresponding to the
            vertical arrows in Fig. 2. The dashed line is the cooling flow 
            model at $t=15$ Gyr. Note the very centrally bright, cuspy profile
            of the cooling flow, the more normal core-halo structure of the
            pre-burst profile, and the low $\SX$ of the post-burst
            profile. \label{fig4}}

\figcaption[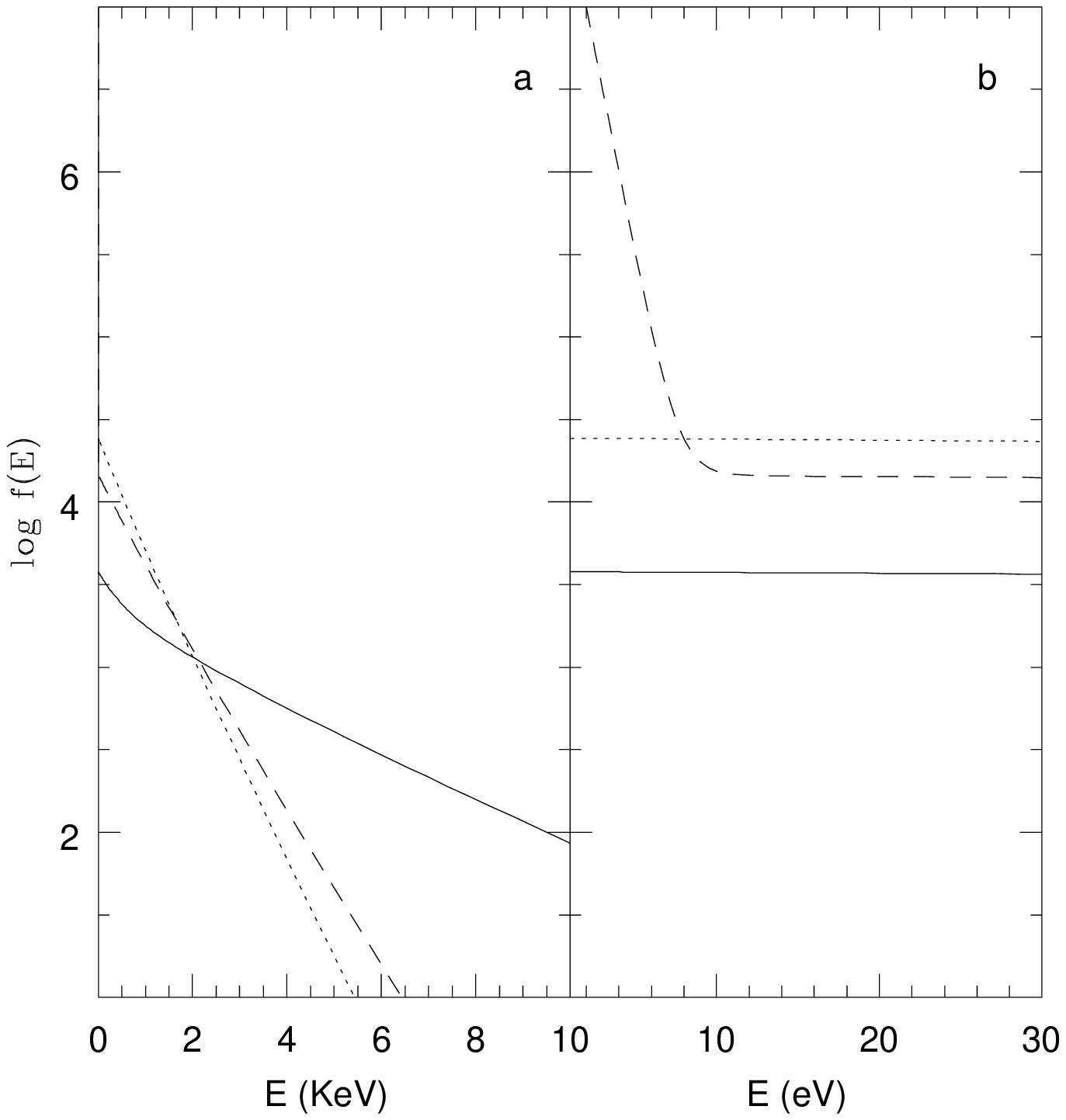]
            {Panel (a): the spectral energy distribution (in arbitrary units)
             of the coronal $\lx$, for the same models shown in Fig. 4.
             Panel (b): blow-up of the previous panel at low energy. Note
             the soft (UV) component in the cooling flow spectrum and the
             hard, high energy tail in the post burst spectrum. \label{fig5}}

%%%%%%%%%%%%%%%%%%%%%%%%%%%%%%%%%%%%%%%%%%%%%%%%%%%%%%%%%%%%%%%%%%%%%

\end{document}